# The effects of four-wheel steering on the path-tracking control of automated vehicles

Sungjin Lim[1], Illés Vörös[2], Yongseob Lim[1*] and Gábor Orosz[2,3*]
[1]*Department of Robotics and Mechatronics Engineering, Daegu Gyeongbuk Institute of Science and Technology, Daegu 42988, South Korea*
[2]*Department of Mechanical Engineering, University of Michigan, Ann Arbor, MI 48109, USA*
[3]*Department of Civil and Environmental Engineering, University of Michigan, Ann Arbor, MI 48109, USA*

**Abstract**

In this study, we analyze the stability of a path-tracking controller designed for a four-wheel steering vehicle, incorporating the effects of the reference path curvature. By employing a simplified kinematic model of the vehicle with steerable front and rear wheels, we derive analytical expressions for the stability regions and optimal control gains specific to different four-wheel steering strategies. To simplify our calculations, we keep the rear steering angle $\delta_\mathrm{r}$ proportional to the front steering angle $\delta_\mathrm{f}$ by using the constant parameter $a$, i.e., $\delta_\mathrm{r} = a\delta_\mathrm{f}$, where $\delta_\mathrm{f}$ is calculated from a control law having both feedforward and feedback terms.

Our findings, supported by stability charts and numerical simulations, indicate that for high velocities and paths of small curvatures, the appropriately tuned four-wheel steering controller significantly reduces lateral acceleration and enhances path-tracking performance when compared to using only front-wheel steering. Furthermore, for low velocities and large curvatures, the using negative $a$ values (i.e., steering the rear wheels in the opposite direction than the front wheels) allows for a reduced turning radius, increasing the vehicle's capability to perform sharp turns in confined spaces like in parking lots or on narrow roads.

* Corresponding author, E-mail: yslim73@dgist.ac.kr, orosz@umich.edu.

## 1 Introduction

The automotive industry continues to enhance safety and driving performance through ongoing technological innovations. Among these advancements, the four-wheel steering (4WS) system has emerged as a pivotal technology that can significantly improve vehicles' handling capabilities [1]. By controlling the steering angles of both the front and rear wheels, the 4WS system enhances vehicle stability, steering responsiveness and maneuverability at both low and high speeds [2]-[4]. This system may be particularly beneficial for improving stability during high-speed driving and for enhancing maneuverability during low speed driving [5], [6], making it versatile across various driving conditions. This contributes to the safety of both the driver and passengers, while it can also reduce the vehicle's turning radius, making parking and maneuvering in tight spaces easier, which is an advantage in urban environments. The purpose of this study is to analyze the key features of 4WS and highlight the performance improvements they offer for automotive systems.

## 2 Vehicle model

In this section, a kinematic bicycle model of a vehicle is derived, considering both the front and rear steering angles. We assume constant longitudinal velocity $V$ and assigned steering angles for both the front and rear wheels, while the roll, pitch, vertical and tire dynamics are neglected. The vehicle parameters contain the wheelbase $f$, the distance $d$ between the center of gravity G and the rear axle center point R, as shown in Fig. 1.

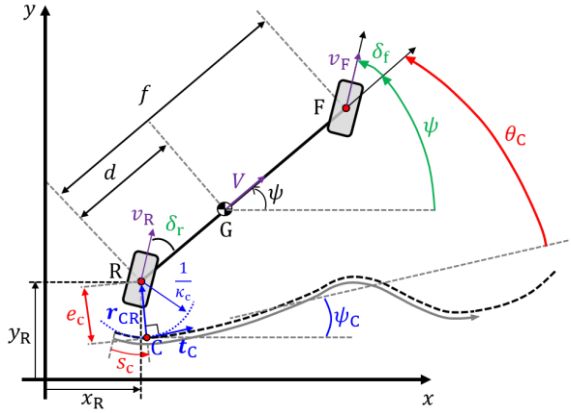

Figure 1: Schematic of a kinematic bicycle vehicle model with four-wheel steering.

### 2.1 Vehicle model in the global reference frame

In the global reference frame, the coordinates $x_R$ and $y_R$ of point R, along with the yaw angle $\psi$, are used as generalized coordinates to describe the vehicle's position and orientation. Since the tire dynamics are neglected, tire slip does not occur and the directions of the velocity vectors at the wheel center points F and R align with the directions of the wheels. The three kinematic constraints of the 4WS vehicle model can be expressed as

$$\begin{cases} \mathbf{v}_F \times \boldsymbol{\rho}_1 = \mathbf{0}, \\ \mathbf{v}_R \times \boldsymbol{\rho}_2 = \mathbf{0}, \\ \mathbf{v}_R \cdot \boldsymbol{\rho}_2 = V, \end{cases} \quad (1)$$

where

$$\mathbf{v}_F = \mathbf{v}_R + \boldsymbol{\omega} \times \mathbf{r}_{RF}, \quad (2)$$

$$\boldsymbol{\rho}_1 = \begin{bmatrix} \cos(\psi + \delta_f) \\ \sin(\psi + \delta_f) \\ 0 \end{bmatrix}, \boldsymbol{\rho}_2 = \begin{bmatrix} \cos(\psi + \delta_r) \\ \sin(\psi + \delta_r) \\ 0 \end{bmatrix}, \quad (3)$$

$$\mathbf{v}_R = \begin{bmatrix} \dot{x}_R \\ \dot{y}_R \\ 0 \end{bmatrix}, \mathbf{v}_F = \begin{bmatrix} \dot{x}_R - f\dot{\psi}\sin\psi \\ \dot{y}_R + f\dot{\psi}\cos\psi \\ 0 \end{bmatrix}. \quad (4)$$

Here $\mathbf{v}_F$ and $\mathbf{v}_R$ are the velocity vectors at points F and R, respectively, while $\boldsymbol{\rho}_1$ and $\boldsymbol{\rho}_2$ are the direction vectors of the front and rear wheels, respectively. Also, $\mathbf{r}_{RF} = [f\cos\psi \quad f\sin\psi \quad 0]^T$ is the vector pointing from R to F, while $\boldsymbol{\omega} = [0 \quad 0 \quad \dot{\psi}]^T$ is the angular velocity vector. Then the constraint equations can be simplified to

$$(\dot{x}_R - f\dot{\psi}\sin\psi)\sin(\psi + \delta_f) - (\dot{y}_R + f\dot{\psi}\cos\psi)\cos(\psi + \delta_f) = 0, \quad (5)$$

$$\dot{x}_R \sin(\psi + \delta_r) - \dot{y}_R \cos(\psi + \delta_r) = 0, \quad (6)$$

$$\dot{x}_R \cos(\psi + \delta_r) + \dot{y}_R \sin(\psi + \delta_r) - V = 0. \quad (7)$$

From these equations the time derivatives of the generalized coordinates can be expressed as

$$\dot{x}_R = V\cos(\psi + \delta_r), \quad (8)$$
$$\dot{y}_R = V\sin(\psi + \delta_r), \quad (9)$$
$$\dot{\psi} = \frac{V\sin(\delta_f - \delta_r)}{f\cos\delta_f}. \quad (10)$$

The velocity of the center of gravity G can be expressed using the transport formula:

$$\mathbf{v}_G = \mathbf{v}_R + \boldsymbol{\omega} \times \mathbf{r}_{RG}, \quad (11)$$

where $\mathbf{r}_{RG} = [d\cos\psi \quad d\sin\psi \quad 0]^T$ is the vector pointing from R to point G. By differentiating equation (11), the acceleration vector of the center of mass G can be calculated as

$$\boldsymbol{a}_G = \begin{bmatrix} a_{Gx} \\ a_{Gy} \\ 0 \end{bmatrix} = \quad (12)$$

$$\begin{bmatrix} -V(\dot{\psi} + \dot{\delta}_r)\sin(\psi + \delta_r) - d\dot{\psi}^2\cos\psi - d\ddot{\psi}\sin\psi \\ V(\dot{\psi} + \dot{\delta}_r)\cos(\psi + \delta_r) - d\dot{\psi}^2\sin\psi + d\ddot{\psi}\cos\psi \\ 0 \end{bmatrix}.$$

Therefore, the lateral acceleration at point G is

$$a_G^{lat} = -a_{Gx}\sin\psi + a_{Gy}\cos\psi \quad (13)$$
$$= V(\dot{\psi} + \dot{\delta}_r)\cos\delta_r + d\ddot{\psi}.$$

### 2.2 Transformation to the path-reference frame

To design the path-tracking controller, the absolute position and orientation $(x_R, y_R, \psi)$ expressed in the Earth-fixed frame are converted to the relative position and orientation $(s_C, e_C, \theta_C)$ with respect to the path, where point C is the closest point of the reference path to point R (see Fig. 1), $s_C$ is the arclength coordinate, while $e_C$ and $\theta_C$ are the lateral and angle errors at point C. The coordinate transformation for an arbitrary point in differential form can be expressed as in [7]:

$$\dot{s}_C = \frac{\dot{x}_R \cos\psi_C + \dot{y}_R \sin\psi_C}{1 - \kappa_C e_C}, \quad (14)$$

$$\dot{e}_C = -\dot{x}_R \sin\psi_C + \dot{y}_R \cos\psi_C, \quad (15)$$

$$\dot{\theta}_C = \kappa_C \frac{\dot{x}_R \cos\psi_C + \dot{y}_R \sin\psi_C}{1 - \kappa_C e_C} + \dot{\psi}, \quad (16)$$

where the angle $\psi_C$ represents the direction of the tangential vector $\mathbf{t}_C$, we used $\theta_C = \psi - \psi_C$, while $\kappa_C$ denotes the curvature of the path at point C. By substituting equations (8), (9), and (10) into equations (14), (15), and (16), the transformed vehicle model becomes

$$\dot{s}_C = \frac{\cos(\theta_C + \delta_r)}{1 - \kappa_C e_C}, \quad (17)$$

$$\dot{e}_C = V \sin(\theta_C + \delta_r), \quad (18)$$

$$\dot{\theta}_C = -\frac{\kappa_C V \cos(\theta_C + \delta_r)}{1 - \kappa_C e_C} + \frac{V \sin(\delta_f - \delta_r)}{f \cos\delta_f}, \quad (19)$$

where the first equation describes the longitudinal motion of point C along the path, while the last two equations give the evolution of lateral deviation and relative yaw angle with respect to the path. In the following, the transformed equations will be used to design a path-tracking controller.

## 3 Path-tracking controller design

In this section, path-tracking controllers are designed to ensure that the point R follows the desired path on the road. In order to simplify the analysis, a straight reference path along the *x*-axis is considered first, using equations (8), (9), and (10). Next, a controller for a reference path with varying curvature is designed using equations (17), (18), and (19).

### 3.1 Straight path-tracking controller

Our path-tracking controller uses a simple linear feedback control law to determine the steering angles for the front and rear wheels, guiding the rear axle center point R along a specified path while maintaining a zero relative yaw angle. The control law is as follows:

$$\delta_f^{FB} = -k_1 y_R - k_2 \psi, \quad (20)$$
$$\delta_r^{FB} = -k_3 y_R - k_4 \psi, \quad (21)$$

and the resulting closed-loop lateral dynamics are

$$\dot{y}_R = V(\psi - k_3 y_R - k_4 \psi), \quad (22)$$
$$\dot{\psi} = \frac{V}{f}\big((-k_1 + k_3)y_R + (-k_2 + k_4)\psi\big), \quad (23)$$

where $k_1$ and $k_2$ are the feedback gains for the front steering angle, $k_3 = ak_1$ and $k_4 = ak_2$ are feedback gains for the rear steering angle, and $a$ is a proportional parameter. Note that if $a = 0$, the system reduces to a front wheel steering vehicle.

### 3.2 Curved path-tracking controller

In this section, both feedback and feedforward controllers are designed to ensure that the rear axle center point R follows the desired path on the road. The feedforward controller can accurately predict the steering angle for a given curvature in steady-state conditions but cannot correct errors caused by the initial state or disturbances. The feedback controller adjusts the steering angle based on real-time measurements to correct these errors. By combining these two control strategies, precise path tracking is achieved, enhancing the overall stability and performance of the vehicle. Based on this, we chose

$$\delta_f = \delta_f^{FF} + \delta_f^{FB}, \quad (24)$$
$$\delta_r = \delta_r^{FF} + \delta_r^{FB}, \quad (25)$$

where $\delta_f^{FF}$ and $\delta_f^{FB}$ are the front feedforward and feedback steering angles, while $\delta_r^{FF}$ and $\delta_r^{FB}$ are the rear feedforward and feedback steering angles. When the lateral deviation and the relative yaw angle become zero, point R moves on the path, and the feedforward term can be determined using equations (18) and (19) yielding

$$\delta_f^{FF} = \tan^{-1}(\kappa_C f), \quad (26)$$
$$\delta_r^{FF} = 0. \quad (27)$$

The feedback controller is designed using proportional terms for the error terms $e_C$ and $\theta_C$, similarly to equations (20) and (21):

$$\delta_f^{FB} = -k_1 e_C - k_2 \theta_C, \quad (28)$$
$$\delta_r^{FB} = -k_3 e_C - k_4 \theta_C, \quad (29)$$

where $k_1$ and $k_2$ are feedback gains for the front steering angle, while $k_3 = ak_1$ and $k_4 = ak_2$ are feedback gains for the rear steering angle.

## 4 Stability analysis

In this section, the linear stability of the proposed controller is analyzed. The steady state solutions of $e_C$ and $\theta_C$ are $e_C^* = 0$ and $\theta_C^* = 0$, respectively, which correspond to following the path perfectly.

## 4.1 Stability of the linearized system

By defining the state perturbations

$$\tilde{e}_C = e_C - e_C^*, \quad (30)$$
$$\tilde{\theta}_C = \theta_C - \theta_C^*, \quad (31)$$

we linearize the system about the steady state resulting in

$$\dot{\mathbf{x}}(t) = \mathbf{A}\mathbf{x}(t) + \mathbf{B}\mathbf{u}(t), \quad (32)$$

where the state and input vectors are $\mathbf{x} = [\tilde{e}_C \quad \tilde{\theta}_C]^T$ and $\mathbf{u} = [\delta_f \quad \delta_r]^T$, respectively, and the feedback law can be formulated as $\mathbf{u} = \mathbf{K}\mathbf{x}$. The matrices are

$$\mathbf{A} = \begin{bmatrix} 0 & V \\ -V\kappa_C^2 & 0 \end{bmatrix}, \mathbf{B} = \begin{bmatrix} 0 & V \\ \frac{V}{f} & -\frac{V}{f} \end{bmatrix}, \quad (33)$$
$$\mathbf{K} = \begin{bmatrix} -k_1 & -k_2 \\ -k_3 & -k_4 \end{bmatrix},$$

The characteristic equation of the linear system is

$$\begin{aligned} D(\lambda) &= \det(\lambda \mathbf{I} - \mathbf{A} - \mathbf{B}\mathbf{K}) \\ &= \lambda^2 + \frac{V}{f}(fak_1 + (1-a)k_2)\lambda \\ &\quad + \frac{V^2}{f}\left((1-a)k_1 + (1-ak_2)f\kappa_C^2\right) \\ &= 0. \end{aligned} \quad (34)$$

For a dynamical system to be stable, all characteristic roots must be located in the left half complex plane. In other words, the real parts of all characteristic roots must be negative, i.e., $\Re(\lambda) < 0$ for all $\lambda \in \mathbb{C}$. By applying the Routh-Hurwitz criteria, we obtain the stability conditions

$$\frac{V}{f}(fak_1 + (1-a)k_2) > 0, \quad (35)$$
$$\frac{V^2}{f}\left((1-a)k_1 + (1-ak_2)f\kappa_C^2\right) > 0. \quad (36)$$

From these conditions the stability boundaries in the $(k_1, k_2)$ plane can be calculated as

$$\begin{cases} k_1 = \frac{(ak_2 - 1)f\kappa_C^2}{1-a} \\ k_2 = -\frac{fak_1}{1-a} \end{cases} \quad \text{if } a \neq 1, \quad (37)$$
$$\begin{cases} k_1 = 0 \\ k_2 \in \mathbb{R} \end{cases} \quad \text{if } a = 1. \quad (38)$$

When $a = 1$, the front and rear steering angles are identical, leading to zero yaw error. As a result, the system is independent of the yaw feedback gain $k_2$, allowing $k_2$ to take any real value.

## 4.2 Pole placement design

To achieve desired control performance, we select the control gains using the pole placement method. We consider the case when the characteristic equation has a double real root at $\lambda = \lambda_0$. Correspondingly, the characteristic equation has the desired form $(\lambda - \lambda_0)^2 = \lambda^2 - 2\lambda_0\lambda + \lambda_0^2 = 0$. By comparing this with the coefficients in equation (34) yields

$$-2\lambda_0 = \frac{V}{f}(fak_1 + (1-a)k_2), \quad (39)$$
$$\lambda_0^2 = \frac{V^2}{f}\left((1-a)k_1 + (1-ak_2)f\kappa_C^2\right), \quad (40)$$

which can be solved explicitly for the control gains $k_1$ and $k_2$. This approach ensures that the system exhibits the desired stability and dynamic performance. For a straight road where $\kappa_C = 0$, the the control gains are

$$\begin{cases} k_1 = \frac{f\lambda_0^2}{V^2(1-a)} \\ k_2 = \frac{-\lambda_0 f}{V(1-a)}\left(2 + \frac{\lambda_0 af}{V(1-a)}\right) \end{cases} \quad \text{if } a \neq 1, \quad (41)$$
$$\begin{cases} k_1 = \frac{-2\lambda_0}{f} \\ k_2 \in \mathbb{R} \end{cases} \quad \text{if } a = 1. \quad (42)$$

On the other hand, the control gains for a curved road are

$$\begin{cases} k_1 = K_1 \\ k_2 = K_2 \end{cases} \quad \text{if } a \neq 0, \quad (43)$$
$$\begin{cases} k_1 = f\left(\frac{\lambda_0^2}{V^2} - \kappa_C^2\right) \\ k_2 = \frac{-2\lambda_0 f}{V} \end{cases} \quad \text{if } a = 0, \quad (44)$$

where

$$K_1 = -\frac{2\lambda_0}{Va} \\ + \frac{V^2 af\kappa_C^2 - 2V\lambda_0(1-a) - af\lambda_0^2}{V^2(a^2f^2\kappa_C^2 + (1-a)^2)}\left(1 - \frac{1}{a}\right), \quad (45)$$

$$K_2 = \frac{f(V^2 af\kappa_C^2 - 2V\lambda_0(1-a) - af\lambda_0^2)}{V^2(a^2f^2\kappa_C^2 + (1-a)^2)}. \quad (46)$$

## 4.3 Analysis of stable regions for different vehicle parameters

The stable regions are illustrated for the parameters $f = 2.7$ [m], $d = 1.35$ [m], for both a low velocity of $V = 5$ [m/s] and a high velocity of $V = 20$ [m/s].

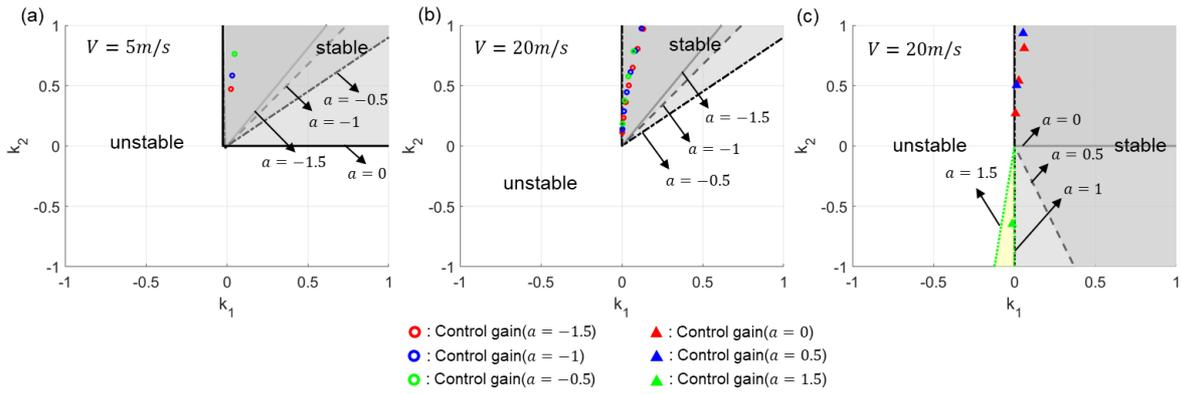

Figure2: The stable domain of control gains for different values of the parameter $a$ and velocity $V$ at zero path curvature $\kappa_C = 0$. (a) for parameters $a = -1.5, -1, -0.5, 0$ and $V = 5$ [m/s], (b) for parameters $a = -1.5, -1, -0.5$ and $V = 20$ [m/s], (c) for parameters $a = 0, 0.5, 1, 1.5$ and $V = 20$ [m/s].

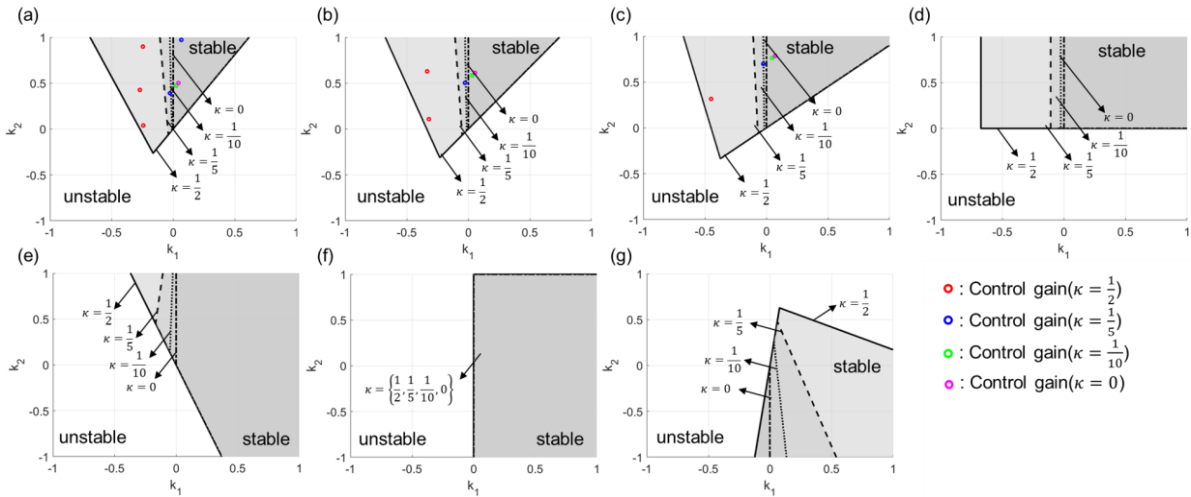

Figure 3: The stable domain of control gains for different values of the parameter $a$ at different path curvature $\kappa_C$ when $V = 5$ [m/s]. (a) for parameter $a = -1.5$, (b) for parameter $a = -1$, (c) for parameter $a = -0.5$, (d) for parameter $a = 0$, (e) for parameter $a = 0.5$, (f) for parameter $a = 1$, (g) for parameter $a = 1.5$.

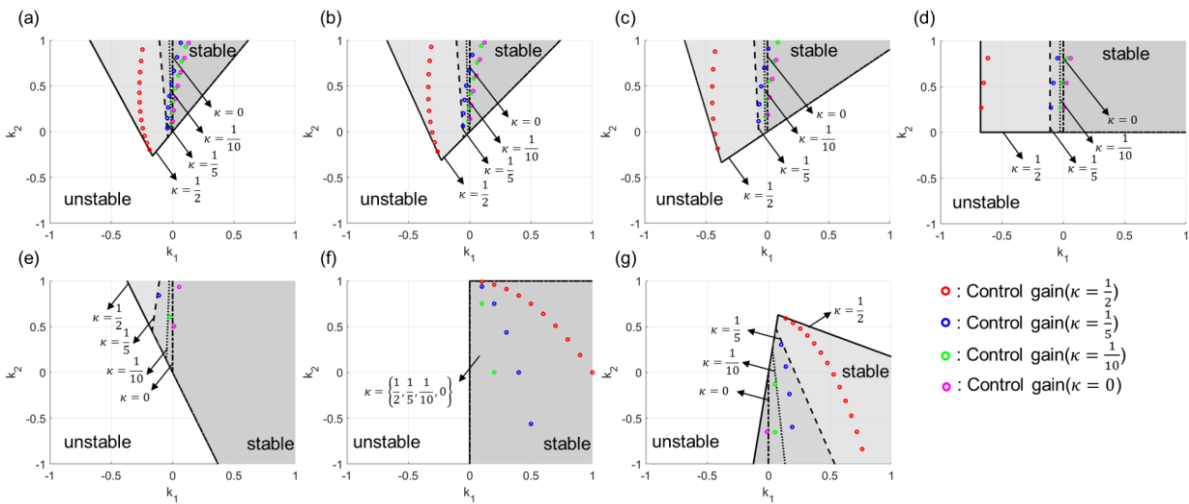

Figure 4: The stable domain of control gains for different values of the parameter $a$ at different path curvatures $\kappa_C$ when $V = 20$ [m/s]. (a) for parameter $a = -1.5$, (b) for parameter $a = -1$, (c) for parameter $a = -0.5$, (d) for parameter $a = 0$, (e) for parameter $a = 0.5$, (f) for parameter $a = 1$, (g) for parameter $a = 1.5$.

Note that the stability boundaries themselves are not influenced by the vehicle's velocity $V$; only the position of the optimal gains is affected, see equations (37)-(38) and (41)-(46).

The stability charts for straight path (i. e., $\kappa_C = 0$) are shown in Fig. 2 for different values of parameters $a$ and $V$. Positive values of $a$ cause the rear steering to rotate in the same direction as the front steering, while negative values of $a$ cause rotations in the opposite direction. As the value of $a$ changes from $-1.5$ to $1.5$, the stable region extends to the right while $k_1 = 0$ remains the left boundary. Note that when $a$ is larger than 1, the stable region extends to the left half plane and it is bounded by the green dashed line.

To avoid the use of large control gains, we consider the case $\lambda_0 = -1$ which can be achieved using gains smaller than 1. In the case of low velocity, this occurs for negative values of $a$, i.e., the rear wheels have to be steered in the opposite direction than the front ones to achieve fast manoeuvring. Note that the case $a = 0$ (front wheel steering) is outside this range, but it is shown for comparison. For non-zero path curvature, the stability charts for different values of parameter $a$ and curvature $\kappa_C$ are shown in Fig. 3 for low velocity and in Fig. 4 for high velocity. The control gains for each value of $a$ are in the range where $k_1$ and $k_2$ are between $-1$ and 1. Note that although the stability boundaries do not depend on $V$, at higher speeds smaller control gains are sufficient to achieve the error decay rate specified by the same $\lambda_0$.

In order to account for passenger comfort, the vehicle's lateral acceleration at point G can be used as an evaluation metric when determining the control gains within the stable region. Fig. 5(a) and (b) show the lateral acceleration for different values of $a$ at low velocity while using different controllers: a feedback controller for the straight road and a feedback+feedforward controller for the curved road of curvature $\kappa_C = 1/10\,[1/m]$. Similarly, Fig. 5(c) and (d) show the lateral acceleration for different values of $a$ at high velocity using the same set of controllers and the curved road of curvature of $\kappa_C = 1/100\,[1/m]$. For the straight road, the initial position is $(x_R(0), y_R(0)) = (0, 2\,m)$. For the curved road at low velocity the initial position is $(x_R(0), y_R(0)) = (0, -5\,m)$, and at high velocity it is $(x_R(0), y_R(0)) = (0, -10\,m)$. The initial angle errors are zeros for all cases. For high velocity the double root values $-1$, $-2$, and $-3$ are used for $\lambda_0$, while for low velocity the values $-1$ and $-2$ are used. Larger (in magnitude) double roots are not considered due to excessive lateral acceleration. The maximum lateral acceleration results show that $\lambda_0 = -1$ yields reasonably matches the actual vehicle's accelerations. Therefore, the rest of the simulations are conducted using the control gains corresponding to this case.

## 5 Numerical simulations

In this section, simulation results using the selected double root $\lambda_0 = -1$ are presented for various values of parameter $a$. These simulations demonstrate scenarios on a straight road using a feedback controller and on a curved road using a feedback+feedforward controller.

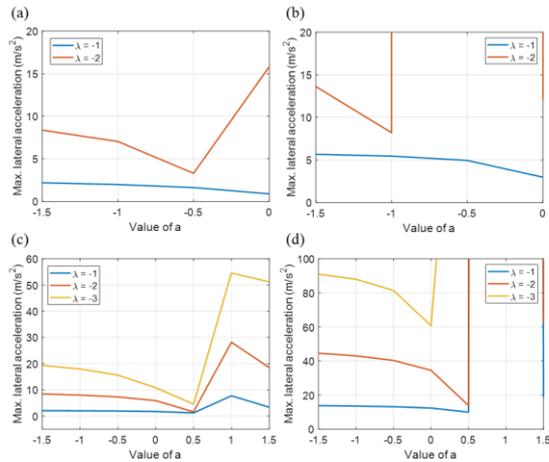

Figure 5: (a)-(b) Simulation results for maximum lateral acceleration for different values of $a$ when $V = 5$ [m/s]. (a) straight road with a feedback controller, (b) curved road with a feedback+feedforward controller. (c)-(d) Simulation results for maximum lateral acceleration for different values of $a$ when $V = 20$ [m/s]. (c) straight road with a feedback controller, (d) curved road with a feedback+feedforward controller.

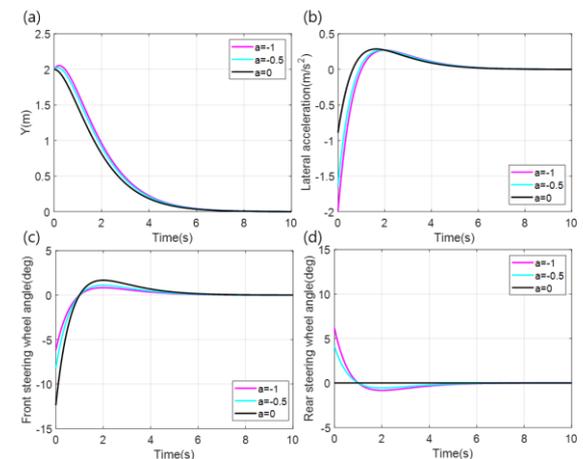

Figure 6: Simulation results for a straight road when $\lambda_0 = -1$ and $V = 5$ [m/s]. (a) lateral position, (b) lateral acceleration, (c) front steering angle, (d) rear steering angle.

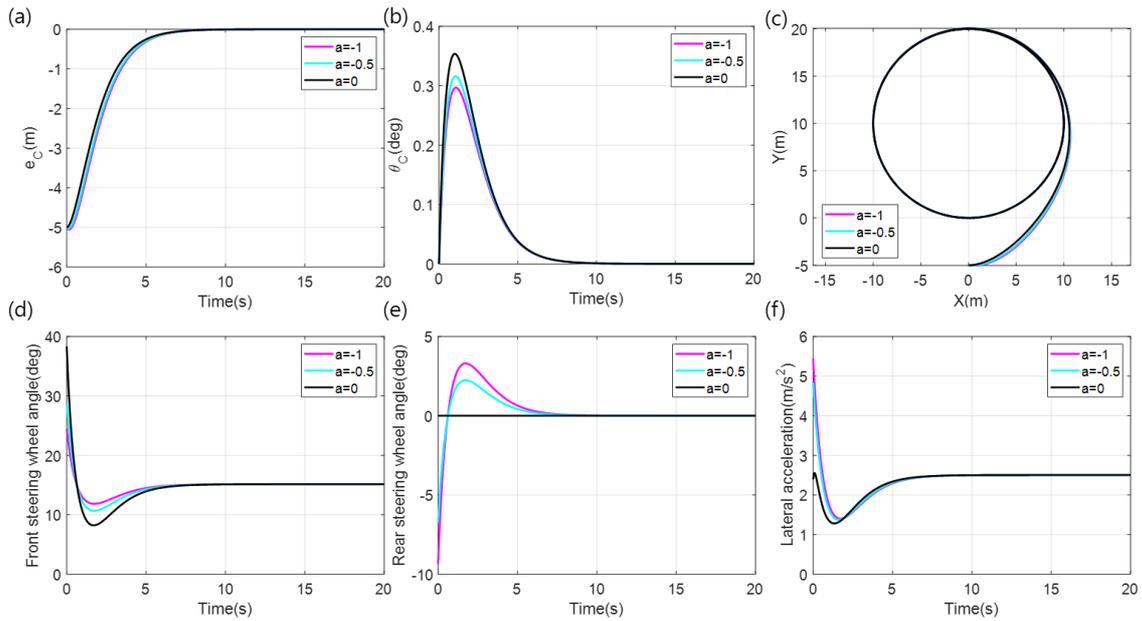

Figure 7: Simulation results for a curved road with a feedback+feedforward controller when $\lambda_0 = -1$ and $V = 5$ [m/s]. (a) lateral error, (b) yaw angle error, (c) vehicle trajectory in the global reference frame, (d) front steering angle, (e) rear steering angle, (f) lateral acceleration.

In this section we only consider cases where the ratio of the rear steering angle to the front steering angle is between $-1$ and $1$. Therefore, only values $a = -1, -0.5, 0, 0.5$ and $1$ are considered.

Figs. 6 and 7 present the simulation results for a straight road and for a curved road of curvature $\kappa_C = 1/10$ [1/m], respectively, when $V = 5$ [m/s]. Fig. 6(a) and (b) present the simulation results for the vehicle's lateral position and lateral acceleration, while Fig. 6(c) and 6(d) present the front and rear steering angles. Fig. 7(a) and Fig. 7(b) present the simulation results of the vehicle's lateral deviation and relative yaw angle, Fig. 7(c) shows the resulting trajectory, Fig. 7(d) and (e) present the front and rear steering angles, and Fig. 7(f) displays the lateral acceleration. Using only a feedback controller for a curved road results in a steady-state error in the lateral deviation and relative yaw angle, leading to degraded path-tracking performance. However, including feedforward control allows the errors to converge to zero, as shown in Fig. 7. For low velocity and large curvature, using a negative $a$ parameter allows for a smaller turning radius, demonstrating the possibility of performing maneuvers such as sharp turns in confined spaces like parking lots and narrow roads.

Figs. 8 and 9 show the simulation results for a straight road and for a curved road with curvature of $\kappa_C = 1/100$ [1/m], respectively, when $V = 20$ [m/s]. For such high velocity and small curvature, using $a = 0.5$ or $a = 1$ show faster convergence of compared to using only front-wheel steering ($a = 0$) on both straight and curved roads. However, for $a = 1$, the acceleration value is greater than for $a = 0$. Therefore, to achieve rapid steering for path tracking, the parameter $a$ can be set close to 1, while for driving focused on ride comfort, it can be set close to 0.5. However, for low velocity and small curvature, although the lateral acceleration or lateral deviation is not smaller compared to a front-wheel steering vehicle, having a negative value for $a$ can result in a smaller turning radius.

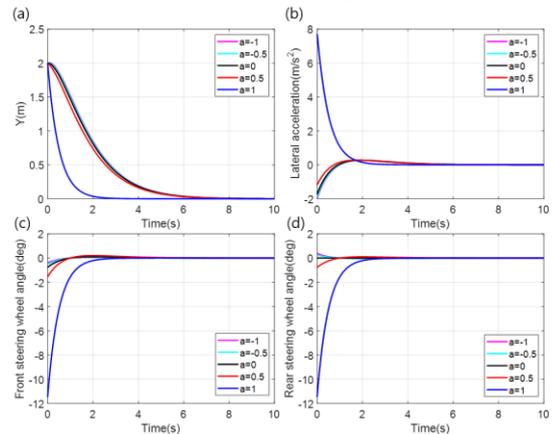

Figure 8: Simulation results for a straight road when $\lambda_0 = -1$ and $V = 20$ [m/s]: (a) lateral position, (b) lateral acceleration, (c) front steering angle, (d) rear steering angle.

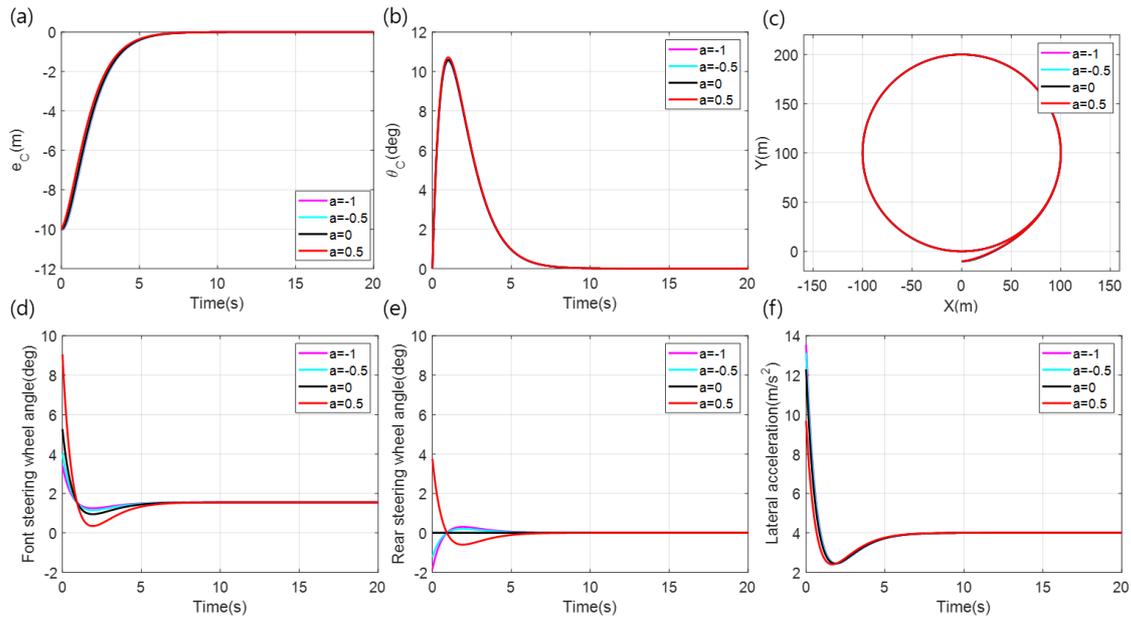

Figure 9: Simulation results for a curved road with a feedback+feedforward controller when $\lambda_0 = -1$ and $V = 20$ [m/s]. (a) lateral error, (b) yaw angle error, (c) vehicle trajectory in the global reference frame, (d) front steering angle, (e) rear steering angle, (f) lateral acceleration.

# 6 Conclusion

In this paper, the stability analysis of a path tracking controller for a four-wheel steering vehicle was performed, with the consideration of the curvature of the reference path. Using a simple kinematic vehicle model, analytical expressions for the stability boundaries and optimal control gains for a linear feedback controller were derived. Numerical simulations show that adjusting the rear-wheel steering controller at high speeds with small curvature improves path-tracking and reduces lateral accelerations. At low speeds with large curvature, it allows for a smaller turning radius, enabling maneuvering in confined spaces.

# Authors

| | |
|---|---|
| 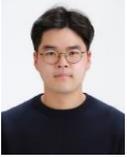 | **SUNGJIN LIM** received his B.S. degree in Mechanical Engineering from Kongju National University, Cheonan, South Korea, and the M.S. degree in Mechanical Engineering from the Inha University, Incheon, South Korea. He is currently pursuing a Ph.D. degree with the Robotics and Mechatronics Engineering Department at the Daegu Gyeongbuk Institute of Science and Technology (DGIST), Daegu, South Korea. His research interests include vehicle dynamics and optimal control, intelligent transportation systems, and their industrial applications. |
| 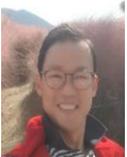 | **YONGSEOB LIM** received his Ph.D. degree in mechanical engineering from the University of Michigan, Ann Arbor, USA, in 2010. He was with the Hyundai Motor Company R&D Advanced Chassis Platform Research Group, South Korea, as a Research Engineer. His work with Hyundai Motor Company focused on advanced vehicle dynamics and control technologies, including active suspension, steering, and braking control systems. Moreover, he was also with Samsung Techwin Company Mechatronics System Development Group, South Korea as a Principal Research Engineer. His work with Samsung focused on developing stabilization control algorithm for the Remotely Controlled Robotic Arms. He is currently an Associate Professor in the Robotics Engineering and Mechatronics Department at the Daegu Gyeongbuk Institute of Science and Technology (DGIST), and the Co-Director of the Joint Research Laboratory of Autonomous Systems and Control (ASC) and Vehicle in the Loop Simulation (VILS) Laboratories. His research interests include the modeling, control, and design of mechatronics systems with interests in autonomous ground vehicles and flight robotics systems and control. |
| 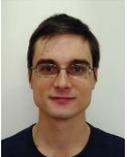 | **ILLÉS VÖRÖS** received the B.Sc. degree in Mechatronics Engineering, and the M.Sc. and Ph.D. degrees in Mechanical Engineering at the Budapest University of Technology and Economics, Hungary, in 2016, 2019 and 2024. He is currently working as a postdoctoral researcher at the University of Michigan, Ann Arbor, USA. His research interests include time delay systems, nonlinear dynamics and control with application to automated vehicles. |
| 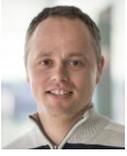 | **GÁBOR OROSZ** received the M.Sc. degree in engineering physics from Budapest University of Technology, Hungary, in 2002, and the Ph.D. degree in engineering mathematics from the University of Bristol, UK, in 2006. He held postdoctoral positions at the University of Exeter, UK, and at the University of California, Santa Barbara, USA. In 2010, he joined the University of Michigan, Ann Arbor, USA, where he is currently a Professor of Mechanical Engineering and a Professor of Civil and Environmental Engineering. From 2017 to 2018, he was a Visiting Professor in Control and Dynamical Systems at the California Institute of Technology, USA. In 2022, he was a Distinguished Guest Researcher in Applied Mechanics at the Budapest University of Technology, where he was a Fulbright Scholar from 2023 to 2024. His research interests include nonlinear dynamics and control, time delay systems, machine learning, and data-driven systems with applications to connected and automated vehicles, traffic flow, and biological networks. |